\documentclass[review]{elsarticle}

\usepackage{lineno,hyperref}
\modulolinenumbers[5]

\journal{Journal of \LaTeX\ Templates}

\usepackage[textsize=footnotesize]{todonotes}

\usepackage{makeidx}
\usepackage{graphics}
\usepackage{color}
\usepackage{listings}
\usepackage{bm}
\usepackage{url}
\usepackage{booktabs}
\usepackage{amsmath}
\usepackage{paralist}
\usepackage{amssymb}
\usepackage{amsthm}
\usepackage{bbold}
\usepackage{subcaption}
\usepackage{algorithm}
\usepackage{algorithmic}
\usepackage{esvect}

\newcommand*\diff{\mathop{}\!\mathrm{d}}

\begin{document}
%=====================================================================================

%=====================================================================================
\begin{frontmatter}

\title{Numerical evidence of electron hydrodynamic whirlpools in graphene samples}

\author{A. Gabbana$^{*}$ }
\address{Universit\`a di Ferrara and INFN-Ferrara, 
         Via Saragat 1, I-44122 Ferrara, Italy. \\
         Bergische Universit\"at Wuppertal, 
         Gau{\ss}strasse 20, D-42119 Wuppertal, Germany. }
\cortext[mycorrespondingauthor]{Corresponding author}
\ead{alessandro.gabbana@unife.it}

\author{M. Mendoza}
\address{ETH Z\"urich, Computational Physics for Engineering Materials, 
         Institute for Building Materials, Schafmattstra{\ss}e 6, HIF, 
         CH-8093 Z\"urich, Switzerland.}

\author{S. Succi}
\address{Istituto per le Applicazioni del Calcolo C.N.R., 
         Via dei Taurini 19, I-00185 Rome, Italy.}

\author{R. Tripiccione}
\address{Universit\`a di Ferrara and INFN-Ferrara, 
         Via Saragat 1, I-44122 Ferrara, Italy.}

%=====================================================================================
\begin{abstract}
%=====================================================================================

We present an extension of recent relativistic Lattice Boltzmann methods
based on Gaussian quadratures for the study of fluids in $(2+1)$ dimensions.
The new method is applied to the analysis of electron flow in graphene samples 
subject to electrostatic drive; we show that the flow displays hydro-electronic whirlpools 
in accordance with recent analytical calculations as well as experimental results.

%=====================================================================================
\end{abstract}
%=====================================================================================


\end{frontmatter}
%=====================================================================================

%\linenumbers

%=====================================================================================
\section{Introduction}
%=====================================================================================

Relativistic fluid dynamics has so far been generally confined to the study 
of astrophysical phenomena, and mostly for ideal non-viscous fluids. 
However, it has been remarked recently that a fluid dynamics approach may be able 
to capture interesting aspects of the behavior of systems at much smaller scales. 
A pioneering example is the study of the behavior of the ultra-relativistic quark-gluon 
plasma formed in the collision of high-energy heavy ions in particle accelerators \cite{heinz-2013, florkowski-2017,romatschke-2017}. 
More recently, it has been suggested that relativistic fluid dynamics is 
relevant to understand the behavior of quantum states that can now be studied in 
several condensed matter experimental setups \cite{fradkin-2010}, within the context of 
so-called AdS-CFT holographic fluids \cite{maldacena-1999,kovtun-2005,bhattacharyya-2008,baier-2008}. 

Recent experimental studies have shown that certain features of the flow of electrons in graphene 
can be explained through a pseudo-relativistic hydrodynamic approach\cite{crossno-2016}, 
confirming earlier theoretical predictions \cite{muller-2008,muller-2009}.

In this paper, we present preliminary results on the development and validation of
computationally efficient numerical approaches aimed at capturing details of the
electron behavior in these systems, addressing the specific case of graphene.

In brief, electrons in graphene follow an ``ultra-relativistic'' dispersion 
relation, so they can be considered as a fluid of massless (quasi-)particles whose 
energy depends on the momentum as $E = v_f p$, with $v_f \sim 10^6$ m/s the 
Fermi speed, mimicking the role of the speed of light in true relativistic systems.
The observation of hydrodynamic regimes is predicted to be simpler in doped
graphene sheets \cite{bandurin-2016}, which are characterized by large viscosities.

The Lattice Boltzmann Method (LBM) is a class of computational fluid 
dynamics solver which has attracted much interest in the past three decades
for the solution of the Navier-Stokes equations. The method stems from the kinetic
level and consists of a quadrature-based discretization of the 
Boltzmann equation, allowing the use of a regular grid of points to 
exactly match the moments of an equilibrium distribution function up
to a desired order. Rigorous mathematical analyses based on the Chapman Enskog expansion
are typically employed to bridge between the kinetic and the macroscopic level
for near equilibrium fluids.
Recently, new formulations have been introduced to extend the method to the 
study of relativistic hydrodynamics. The relativistic Lattice Boltzmann Method (RLBM) \cite{mendoza-2010a,mendoza-2010b}
offers an appealing solution for the study of dissipative relativistic hydrodynamics,
since viscosity is naturally included in its formulation and, furthermore, 
it does not involve second order derivatives in space, preserving 
relativistic invariance and causality by construction \cite{romatschke-2017b}.

The interest towards the numerical study of electrons flow in graphene has
motivated the development of two-dimensional RLBM solvers
\cite{mendoza-2011, oettinger-2013, furtmaier-2015, coelho-2017}. 
Most of these numerical methods are based on a second order expansion of an
equilibrium distribution function following the Fermi-Dirac statistics, and
they have been applied to study e.g. low-viscosity pre-turbulent regimes.

In \cite{gabbana-2017b}, working in three dimensions, we have shown that 
third order expansions of the equilibrium distribution function are the 
minimum requirement to correctly handle dissipative effects in simulations 
of the relativistic regime. 
%
% Models exceeding this minimal requirement do exist,  
% for instance \cite{coelho-2017} supports a fifth order expansion of the equilibrium 
% distribution, but they are not compatible with a Cartesian grid thus requiring interpolation, 
% that severely affects the computational accuracy and efficiency of the corresponding algorithms.

In this paper, we present a new RLBM in two dimensions, based on a third order 
expansion of the equilibrium Maxwell-J\"uttner distribution.
Quantum effects are not described in this model, a choice which simplifies
the algorithmic derivation allowing us to retain one of the  main LBM features, namely perfect streaming.
This could be regarded as the first step in the derivation of a truly accurate model for the fluid dynamics 
descriptions of electrons in graphene, but we also expect that quantum effects should have a 
limited impact on the averages involved in hydrodynamical bulk observables. 
As a result, we expect that the present model should be able to provide new useful insights into the physics 
of relativistic electron flow in graphene devices.
We provide a first validation test of our approach simulating a doped single 
layer graphene sheet in the so-called "vicinity-geometry", which was considered in 
a series of papers \cite{torre-2015, bandurin-2016, pellegrino-2016} 
to outline phenomena such as negative nonlocal resistance and current whirlpools.
The numerical method is tested in a steady-state regime, for which 
semi-analytical solutions are available, showing satisfactory 
agreement with previous works; we wish to emphasize that the present 
numerical method allows to describe time-dependent, nonlinear flows 
which escape analytical treatments. Hence, the future plan is to address 
electron flows of experimental interest.

This paper is organized as follows. In Section~\ref{sec:model} we describe our new RLBM
model, summarizing the general procedure used in its derivation and providing
details of the quadrature and of the external forcing scheme.
In Section~\ref{sec:numerics} we carry out simulations of the "vicinity-geometry"
replicating the formation of current whirlpools and providing quantitative
comparisons of the electrochemical potential against analytical approximations available in the literature.

%=====================================================================================
\section{Model Description}\label{sec:model}
%=====================================================================================
This section briefly describes our RLBM model, which is a "downsizing" to two spatial 
dimensions of a similar model handling relativistic flows in 3D; for a more detailed description, 
the reader is referred to \cite{gabbana-2017a,gabbana-2017b}.

%=====================================================================================
\subsection{Relativistic Boltzmann equation}
%=====================================================================================

We consider an ideal non-degenerate relativistic fluid, consisting at the
kinetic level of a system of interacting particles of rest mass $m$. The particle 
distribution function $f( (x^{\alpha}), (p^{\alpha}) )$, depending on space-time coordinates 
$\left( x^{\alpha} \right) = \left( ct, \bm{x} \right)$ and momenta 
$\left( p^{\alpha} \right) = \left( p^0, \bm{p} \right) = \left( \sqrt{\bm{p}^2 + m^2}, \bm{p} \right)$ 
($c$ is the speed of light, $\bm{x}$, $\bm{p} \in \mathbb{R}^2 $), describes the probability 
of finding a particle with momentum $\bm{p}$ at a given time $t$ and position $\bm{x}$. 
We adopt Einstein's summation convention over repeated indexes, and use Greek indexes 
to denote $(2+1)$ space-time coordinates and Latin indexes for $2$ dimensional spatial coordinates.
The particle distribution function obeys the relativistic Boltzmann equation, here taken in 
the Anderson-Witting \cite{anderson-witting-1974a, anderson-witting-1974b} relaxation-time approximation: 
\begin{equation}\label{eq:relativistic-boltzmann}
  p^{\alpha} \frac{\partial f}{\partial x^{\alpha}} + K^{\alpha} \frac{\partial f}{\partial p^{\alpha}}  
  = \frac{p^{\alpha} U_{\alpha}}{c^2 ~\tau} \left( f - f^{eq} \right) \quad ,
\end{equation}
with $\tau$ the relaxation (proper-)time, $(U^{\alpha}) = \gamma \cdot (c, \bm{u}) $  
the macroscopic $(2+1)$-velocity ($\gamma = 1 / \sqrt{ 1 - \bm{u}^2 / c^2}$), 
$K^{\alpha}$ the external forces acting on the system (for simplicity we assume 
they do not depend on the momentum), and $f^{eq}$ the local equilibrium.
In this work $f^{eq}$ will follow a Maxwell-J\"uttner distribution:
\begin{equation}\label{eq:maxwell-juttner}
  f^{eq} = \frac{1}{A} \exp{ \left( - \frac{p^{\alpha} U_{\alpha}}{k_B T} \right) } \quad ,
\end{equation}
where $A$ is a normalization constant and $k_B$ the Boltzmann constant.

At the macroscopic level the Anderson-Witting model correctly reproduce the conservation
equations, i.e. $\partial_{\alpha} N^{\alpha} = 0$ and $\partial_{\beta } T^{\alpha \beta} = 0$,
with $N^{\alpha}$ the particle $(2+1)$-flow and $T^{\alpha \beta}$ the energy-momentum tensor.
At equilibrium $N^{\alpha}$ and $T^{\alpha \beta}$ can be described by the moments of 
the equilibrium distribution function:
\begin{align}
  N_E^{\alpha             } &= \int f^{eq} p^{\alpha}           \frac{\diff \bm{p}}{p_0}   \label{eq:mj-first-moment}  = n U^{\alpha} \quad , \\
  T_E^{\alpha \beta       } &= \int f^{eq} p^{\alpha} p^{\beta} \frac{\diff \bm{p}}{p_0}   \label{eq:mj-second-moment} = \left( \epsilon + P \right) U^{\alpha} U^{\beta} - P \eta^{\alpha \beta} \quad ,
\end{align}
where $n$ is the particle number-density, $\epsilon$ the energy density, 
$P$ the pressure and $\eta^{\alpha \beta}$ the Minkowski metric tensor.
To be noted that the normalization constant in Eq.~\ref{eq:maxwell-juttner} has to be chosen
in such a way to satisfy the relation with the fluid particle density in Eq.~\ref{eq:mj-first-moment}.
%
%In the following we will use $\eta^{\alpha \beta} = diag(1, -\mathbb{1})$, 
%with $\mathbb{1} = \left( 1, \dots, 1 \right) \in \mathbb{N}^D $,
In the following we will use $\eta^{\alpha \beta} = diag(1, -1, -1)$,
and adopt natural units for which $c = k_B = 1$ .

%=====================================================================================
\subsection{Lattice discretization}
%=====================================================================================

In this section we revise the general procedure, used in the derivation of previous 
non-relativistic \cite{higuera-1989, xiaoyi-luo-1997, shan-1997, martys-1998} 
and relativistic LBMs \cite{romatschke-2011, mendoza-2013, gabbana-2017a}, 
for the discretization of the Boltzmann equation on a lattice.
%
%[un appunto sul fatto che questo approccio permette di recuperare NS..]

We start from an expansion of the equilibrium distribution 
function $f^{eq}$ in a basis of polynomials, orthogonal with respect to 
a weighting function $\omega$  corresponding to $f^{eq}$ in the fluid rest 
frame (where $\bm{u} = 0$). 
It is simple to verify that in the rest frame Eq.~\ref{eq:maxwell-juttner} reduces to 
\begin{equation}\label{eq:mj-poly-weighting-factor}
  \omega(p^0) 
  = 
  \frac{1}{N_R} \exp{\left( -p^0 / T \right)} \quad ,
\end{equation}
where the normalization factor $N_R$ is taken 
such that $\int \omega(p^0) \diff \bm{p} / p^0 = 1$.
Starting from the basis $ \mathcal{V} = \{ 1, p^{\alpha}, p^{\alpha} p^{\beta}, \dots \}$ one
derives the set of orthogonal polynomials $\{J^{(i)}, i = 1,2\dots \}$ by following a  
Gram-Schmidt procedure, with the inner product defined using the weighting function
in Eq.~\ref{eq:mj-poly-weighting-factor}. In \autoref{sec:appendixA} we 
provide an example of polynomials up to the third order for $m=0$.
The polynomials are then used to build the expansion:
\begin{equation}\label{eq:feq-expansion}
  % f^{eq}(p^{\mu}, U^{\mu}, T) 
  % = 
  % \omega( p^0) \sum_{k = 0}^{\infty} a^{(k)}(U^{\mu}, T) J^{(k)} (p^{\mu}) \quad ,
  f^{eq}\left( (p^{\mu}), (U^{\mu}), T \right) 
  = 
  \omega( p^0) \sum_{k = 0}^{\infty} a^{(k)}( (U^{\mu}), T) J^{(k)}( (p^{\mu}) ) \quad ,
\end{equation}
where $a^{(k)}$ are the projection coefficients defined as
\begin{equation}\label{eq:projection-coefficients}
  a^{(k)}( (U^{\mu}) , T) 
  = 
  \int f^{eq}( (p^{\mu}), (U^{\mu}), T)  J^{(k)}( (p^{\mu}) ) \frac{\diff \bm{p}}{p^0}  \quad .
\end{equation}
Observe that by construction the coefficients $a^{(k)}$ coincide with the moments of the 
distribution function; this is a crucial aspect since it follows that $f_N^{eq}$, %$f_N^{eq}( (p^{\mu}), (U^{\mu}), T)$, 
obtained truncating the summation in Eq.~\ref{eq:feq-expansion} such to include only the 
terms of order up to $N$, correctly preserves the moments of the distribution up to 
the $N-th$ order.

The next step consists in determining a Gauss-type quadrature on a Cartesian grid,
with the aim of i) ensuring exact streaming by requiring that all quadrature points 
lie on lattice sites ii) preserving the moments of the distribution up to a desired order $N$.
The discretized version of the equilibrium distribution can be then written as follows:
\begin{equation}\label{eq:feq-expansion-discrete}
  % f_{i N}^{eq}(p^{\mu}, U^{\mu}, T) 
  % = 
  % w_i \sum_{k = 0}^{N} a^{(k)}(U^{\mu}, T) J^{(k)} (p_i^{\mu}) \quad .
  f_{i N}^{eq}( (p^{\mu}), (U^{\mu}), T) 
  = 
  w_i \sum_{k = 0}^{N} a^{(k)}( (U^{\mu}), T) J^{(k)} ( (p_i^{\mu}) ) \quad .
\end{equation}
where $w_i$ and $p_i^{\mu}$ are the weights and the nodes of the quadrature, respectively.
The analytic expression of $f_{i N}^{eq}$ for $N=3$ and $m=0$ is given in \autoref{sec:appendixC}.

At this stage it is possible to formulate the discrete Boltzmann equation, which in the 
relativistic case reads as
\begin{equation}\label{eq:discrete-rbe}
  f_i(\bm{x} + v^{i} \Delta t, t + \Delta t) - f_i(\bm{x}, t) 
  = 
  - \Delta t~ \frac{p_i^{\mu} U_{\mu}}{p^0 \tau} (f_i - f_{i N}^{eq}) + F^{ext}_i  \quad .
\end{equation}

A detailed description of the algorithmic derivation for the 3-dimensional
case is given in \cite{gabbana-2017a}. The algebraic complexities in the 
calculation of the polynomials and the expansion of the equilibrium distribution 
significantly simplify in 2-D. The full details will be 
described at length in a future expanded version of this work.

\subsection{Quadrature with prescribed nodes}

As discussed in the introduction, here we focus our attention on solving 
Eq.~\ref{eq:orthonormal} using polynomials up to the third order.

The lattice discretization of the
Boltzmann equation can be reduced to a quadrature problem. In practice, one needs to
find the weights  and the abscissas of a quadrature able to satisfy the
orthonormal conditions  up to a desired order \cite{philippi-2006}:
\begin{equation}\label{eq:orthonormal}
  % \int \omega(p^0) J_l(p^{\mu}) J_k(p^{\mu}) \frac{\diff \bm{p}}{p^0} 
  % = 
  % \sum_i w_i J_l(p^{\mu}_{i})J_k(p^{\mu}_{i}) = \delta_{lk} \quad ,
  \int \omega(p^0) J_l( (p^{\mu}) ) J_k( (p^{\mu}) ) \frac{\diff \bm{p}}{p^0} 
  = 
  \sum_i w_i J_l( (p^{\mu}_{i})) J_k( (p^{\mu}_{i}))  = \delta_{lk} \quad ;
\end{equation}
here $p^\mu_{i}$ are the discrete quadri-momentum vectors.
A convenient parametrization of $p^\mu_{i}$ was given in \cite{gabbana-2017a}
and  writes as follows:
\begin{equation}\label{eq:discrete-quadri-momentum}
  (p^\mu_{i}) = p^0_i (1, v_0 n_{i}) \quad ,
\end{equation}
where $n_{i} \in \mathbb{Z}^2$ are the vectors forming the stencil $G = \{ n_i ~|~ i = 1,2,\dots,i_{max} \}$ 
defined by the (on-lattice) quadrature points, $v_0$ is a free parameter that can be freely chosen such 
that $v_i = v_0 || n_i || \leq 1, \forall i$, and $p_i^0$ is defined as
\begin{equation}\label{eq:discrete-p0}
  p_i^0 = m \gamma_i = m \frac{1}{\sqrt{1 -v_i^2}} \quad .
\end{equation}
In order to determine a quadrature we proceed as follows: i) select a value for the rest mass
$\bar{m} = m / T_0$ (with $T_0$ a reference temperature on the lattice), 
ii) choose a set of velocity vectors $G$, formed by a sufficient number
of elements such that the left hand side of Eq.~\ref{eq:orthonormal} is a full ranked matrix,
iii) look for a solution of Eq.~\ref{eq:orthonormal} formed by non-negative weights 
$(w_i \ge 0, \forall i)$.

We point out that the parametrization in Eq.~\ref{eq:discrete-quadri-momentum} is general 
and can be used to determine quadratures for wide ranges of values of $\bar{m}$.

\begin{figure}[htb]
  \begin{subfigure}{0.5\textwidth}
    \includegraphics[width=\linewidth]{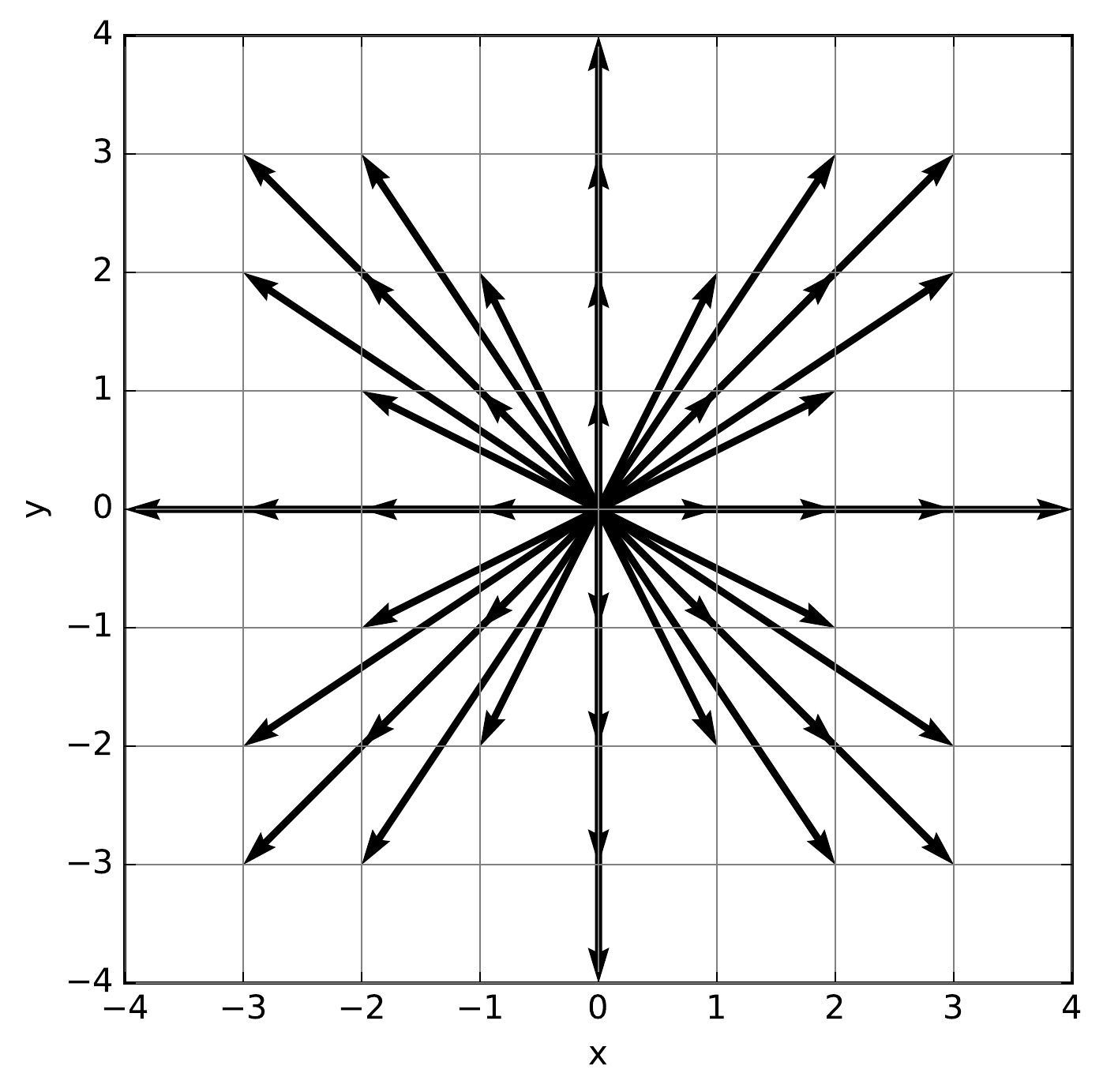}
    \caption{ $\bar{m} = 5$, order 3 } \label{fig:m5-stencil}
  \end{subfigure}
  \hspace*{\fill} % separation between the subfigures
  \begin{subfigure}{0.5\textwidth} 
    \includegraphics[width=\linewidth]{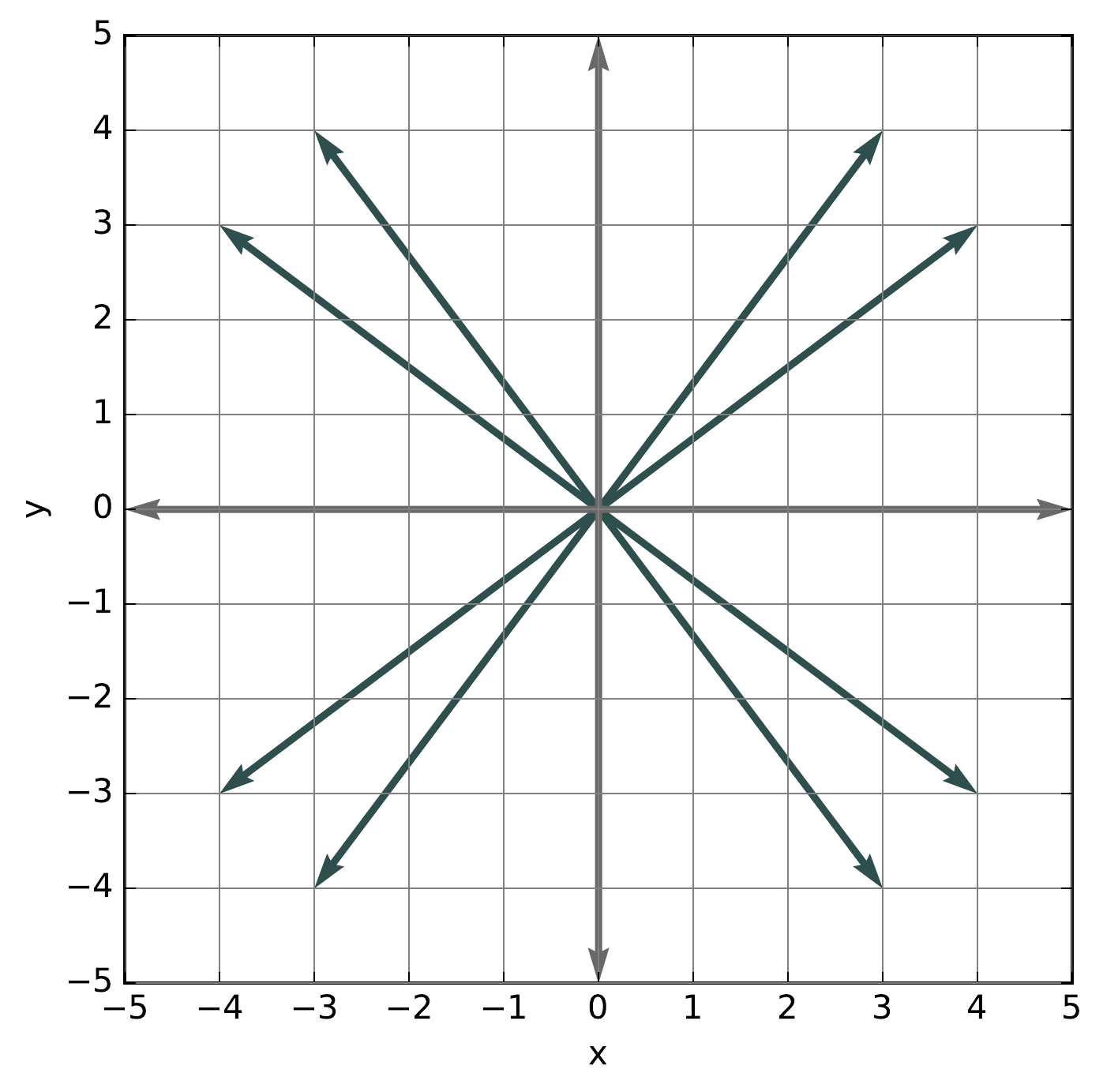}
    \caption{ $\bar{m} = 0$, order 3 } \label{fig:m0-o3}
  \end{subfigure}
  \caption{Examples of stencils for a third-order approximation. 
           Left: $\bar{m} = 5$.
           $G       = \{ \left(     0,     0 \right)     ,$
           $             \left( \pm 1,     0 \right)_{FS},$
           $             \left( \pm 1, \pm 1 \right)_{FS},$
           $             \left( \pm 2,     0 \right)_{FS},$
           $             \left( \pm 2, \pm 1 \right)_{FS},$
           $             \left( \pm 2, \pm 2 \right)_{FS},$
           $             \left( \pm 3,     0 \right)_{FS},$
           $             \left( \pm 3, \pm 2 \right)_{FS},$
           $             \left( \pm 3, \pm 3 \right)_{FS},$
           $             \left( \pm 4,     0 \right)_{FS} $
           $          \}$  (45 components).
           Right: $\bar{m} = 0$. 
           $G = \{ \left( \pm 3, \pm 4 \right)_{FS} , \left( \pm 5, 0 \right)_{FS}  \}$ with 4 energy shells (48 components).
          }\label{fig:stencil}
\end{figure}

%We now focus on the case $D = 2$;
As an example   we show in
\autoref{fig:m5-stencil} a set of vectors that can be used to build a quadrature
for $\bar{m} = 5$.
In the remainder of this paper we are interested in particular in the case of
massless particles, all traveling at the same speed $v_i = c = 1, \forall i$. 
Since for $m = 0$ Eq.~\ref{eq:discrete-p0} is not well defined, we let
$p^0_i$ be free parameters (as already suggested in \cite{mendoza-2013}) to be
determined such as to satisfy Eq.~\ref{eq:orthonormal}.  We can have several
energy shells associated to each vector and therefore we add a second index to
Eq.~\ref{eq:discrete-quadri-momentum}:
\begin{equation}
  (p^\mu_{i,j}) = p^0_j (1, \frac{n_i}{||n_i||}) \quad ,
\end{equation}
where the index $j$ labels different energy shells.

The minimal stencil structure, supporting a third order expansion of the equilibrium distribution function,
has radius $R = ||n_i|| = 5$ (\autoref{fig:m0-o3}); it is formed by the following set of velocity vectors
$G = \{ \left( \pm 3, \pm 4 \right)_{FS} , \left( \pm 5, 0 \right)_{FS}  \}$ (FS stands for
full symmetric), with four energy shells and the following weights:
\begin{align*}
  w_{11} &= 0.003930503244 \dots ~~~~ w_{21} = 0.054642060984 \dots ~~~~ p^0_1 = 0.000016359462 \dots \\
  w_{12} &= 0.008026424774 \dots ~~~~ w_{22} = 0.013535762740 \dots ~~~~ p^0_2 = 3.305423649330 \dots \\
  w_{13} &= 0.000175706060 \dots ~~~~ w_{23} = 0.000296310700 \dots ~~~~ p^0_3 = 7.758786843141 \dots \\
  w_{14} &= 0.042659667266 \dots ~~~~ w_{24} = 0.071941262878 \dots ~~~~ p^0_4 = 0.935838587521 \dots \\
\end{align*}
$w_{1 j}$ and $w_{2 j}$, $j=1,\dots,4$ are respectively the weights associated to the
stencil components $\left( \pm 3, \pm 4 \right)_{FS}$ and $\left( \pm 5, 0 \right)_{FS}$.
This lattice will be used in Section~\ref{sec:numerics} for the numerical part
of this work; in \autoref{sec:appendixA} and \autoref{sec:appendixB} we list
the polynomials and the projections used for the derivation of the method. 

%=====================================================================================
\subsection{Forcing Scheme}
%=====================================================================================
%
The definition of force in relativity is subject to a certain degree of arbitrariness
due to the lack of certain general properties such as, for example, Newton's third law \cite{tuval-2014}. 
In the following we will use the definition of the Minkowski force:
\begin{equation}\label{eq:minkowski-force}
  K^{\alpha} = m \frac{\diff U^{\alpha}}{\diff \tau} \quad ,
\end{equation}
subject to the condition 
\begin{equation}
  K^{\alpha} p_{\alpha} = K^0 p^0 - \bm{K} \cdot \bm{p} = 0 \quad ,
\end{equation}
and
\begin{equation}\label{eq:rel-force}
  \bm{K} = \gamma ~ \bm{F}  \quad .
\end{equation}

To introduce a forcing term in our numerical scheme we make the following two 
assumptions: i) the force does not depend on the momentum three vector 
( $\frac{\partial K^{\alpha}}{\partial p^{\alpha}} = 0  $ ) ii) the distribution function
in not far from the equilibrium, such that we can approximate the term 
$K^{\alpha} \frac{\partial f}{\partial p^{\alpha}}$ in Eq.~\ref{eq:relativistic-boltzmann}
with an expansion that uses the same polynomials used for the equilibrium distribution
function:
\begin{equation}
  \frac{\partial f}{\partial p^{\alpha}} \approx \frac{\partial f^{eq}}{\partial p^{\alpha}} 
  = 
  \omega( p^0) \sum_{k = 0}^{\infty} b^{(k)}( (U^{\mu}), T) J^{(k)} ( (p^{\mu}) )
\end{equation}
with the projection coefficients defined as
\begin{equation}
  b^{(k)}( (U^{\mu}), T) = \int \frac{\partial }{\partial p^{\alpha}}f^{eq}( (p^{\mu}), (U^{\mu}), T)  J^{(k)}( (p^{\mu})) \frac{\diff \bm{p}}{p^0}  \quad .
\end{equation}

% %=====================================================================================
% \subsection{Lattice Boltzmann Algorithm}
% %=====================================================================================
%
% For each time step the following operations are performed on each grid site:

% \begin{enumerate} 
%   \item Compute the first and second moment of the distribution:
%         $$ N^{\alpha}       = \sum_i f_i p^{\alpha}_i$$
%         $$ T^{\alpha \beta} = \sum_i f_i p^{\alpha}_i p^{\beta}_i$$
%   \item Compute the energy density $\epsilon$ and the four velocity $U^{\alpha}$ solving
%         $$ \epsilon U^{\alpha} = T^{\alpha \beta} U_{\beta} $$
%   \item Compute the particle density from
%         $$ n = U_{\alpha} N^{\alpha} $$
%   \item Compute the temperature from the EOS (Eq.~\ref{eq:eos-mass}).

%   \item Compute the equilibrium distribution function:
%         $$ f_i^{eq}(\bm{p},U^{\mu}, T) = w_i \sum_{k = 0}^{N} a^{(k)}(U^{\mu}, T) J^{(k)} (\bm{p}) \quad , $$

%   \item Compute the Minkowsky forcing term

%   \item Evolve the system via the discrete Boltzmann equation:
%         $$ f_i(\bm{x} + v^{i} \delta t, t + \delta t) - f_i(\bm{x}, t) = - \delta t~ \frac{p_i^{\mu} U_{\mu}}{p^0 \tau} (f_i - f_i^{eq}) + K_i $$
% \end{enumerate}

%=====================================================================================
\section{Numerical Tests}\label{sec:numerics}
%=====================================================================================

We now apply the model described in the previous section to the simulation of the 
(pseudo)-relativistic dynamics of electrons in graphene sheets; as already remarked, 
in this case the Fermi velocity $v_f$ of the simulated system plays the role of the speed of light.
We consider an experimental setup consisting of an ultraclean single
layer graphene encapsulated between boron nitride crystals in which it has been
shown that electrons exhibit a hydrodynamic flow \cite{bandurin-2016}. 
This setup has been used in a series of works \cite{torre-2015,bandurin-2016,pellegrino-2016} 
to highlight peculiar properties such as negative nonlocal resistance and current whirlpools.
\begin{figure}[htb]
  \includegraphics[width=\linewidth]{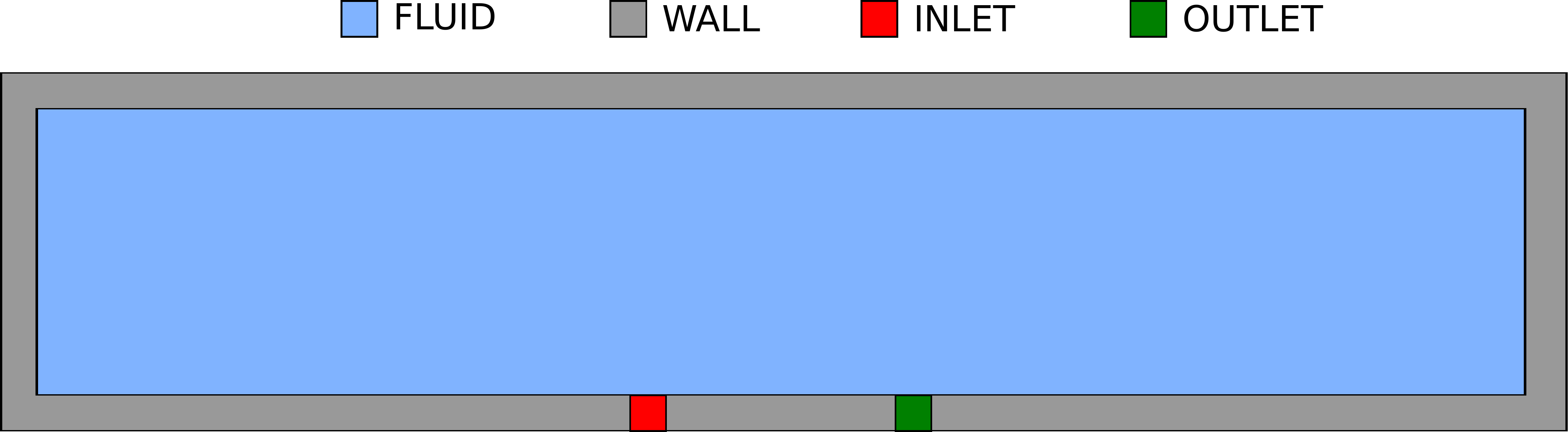}
  \caption{Geometry used for the validation of the code.
           Bounce back boundary conditions are imposed at the wall.
           Sites representing the inlet and the outlet do not evolve in time.
          }\label{fig:geom}
\end{figure}
The geometry is sketched in \autoref{fig:geom}. The total force acting on the system
is given by the vector sum of the force due to the electric field $F_{E}$ and the force
due to the pressure gradient $F_{P}$. While $F_{P}$ is naturally described by our RLBM solver,
$F_{E}$ is included in the form of an external force. Therefore, 
in the simulations the external force $\bm{F}$ (Eq.~\ref{eq:rel-force}) is given by
a self-consistent electric field $\bm{E} = - \rho_e \nabla \phi$, with $\rho_e = n e $ 
being the electron charge density.
For our initial validation tests on this specific setup, we follow \cite{tomadin-2013} 
and do not solve explicitly the Poisson equation for the electric potential, but rather use a 
local capacitance approximation defined as:
\begin{equation}\label{eq:lca}
  \phi(\bm{x}) = - e n(\bm{x}) / C_g \quad ,
\end{equation}
where $C_g$ is the capacitance per unit area. 

\begin{figure}[htb]
  \includegraphics[width=\linewidth]{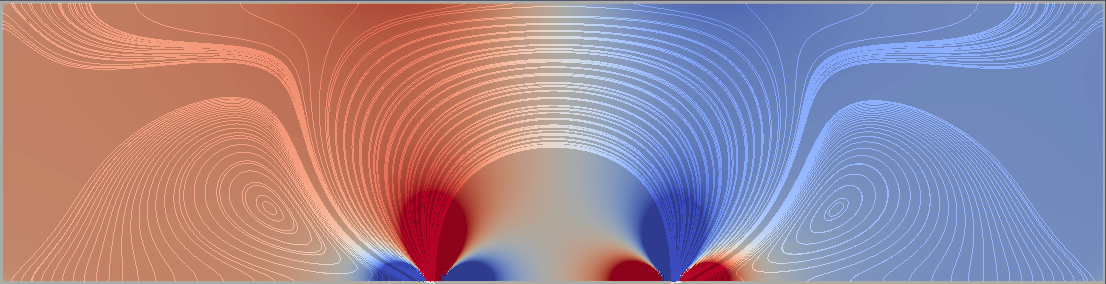}
  \caption{ Snapshot of a simulation on a $2000 \times 500$ lattice, 
            taken after 100000 time steps, with an initial uniform
            density $n = 1.5$, initial $T = 1.25$, a fixed 
            velocity at inlet $v^{in} = 10^{-5}$, $\tau = 1.0$, $C_g = 10$.
            The color map describes the electrochemical potential (red colors positive potential,
            blue colors negative potential). Ticked lines represent 
            the electrons velocity streamlines.
          }\label{fig:whirlpool-rlbm}
\end{figure}

\begin{figure}[htb]
    \centering
    \begin{subfigure}[b]{0.49\textwidth}   
        \centering 
        \includegraphics[width=\textwidth]{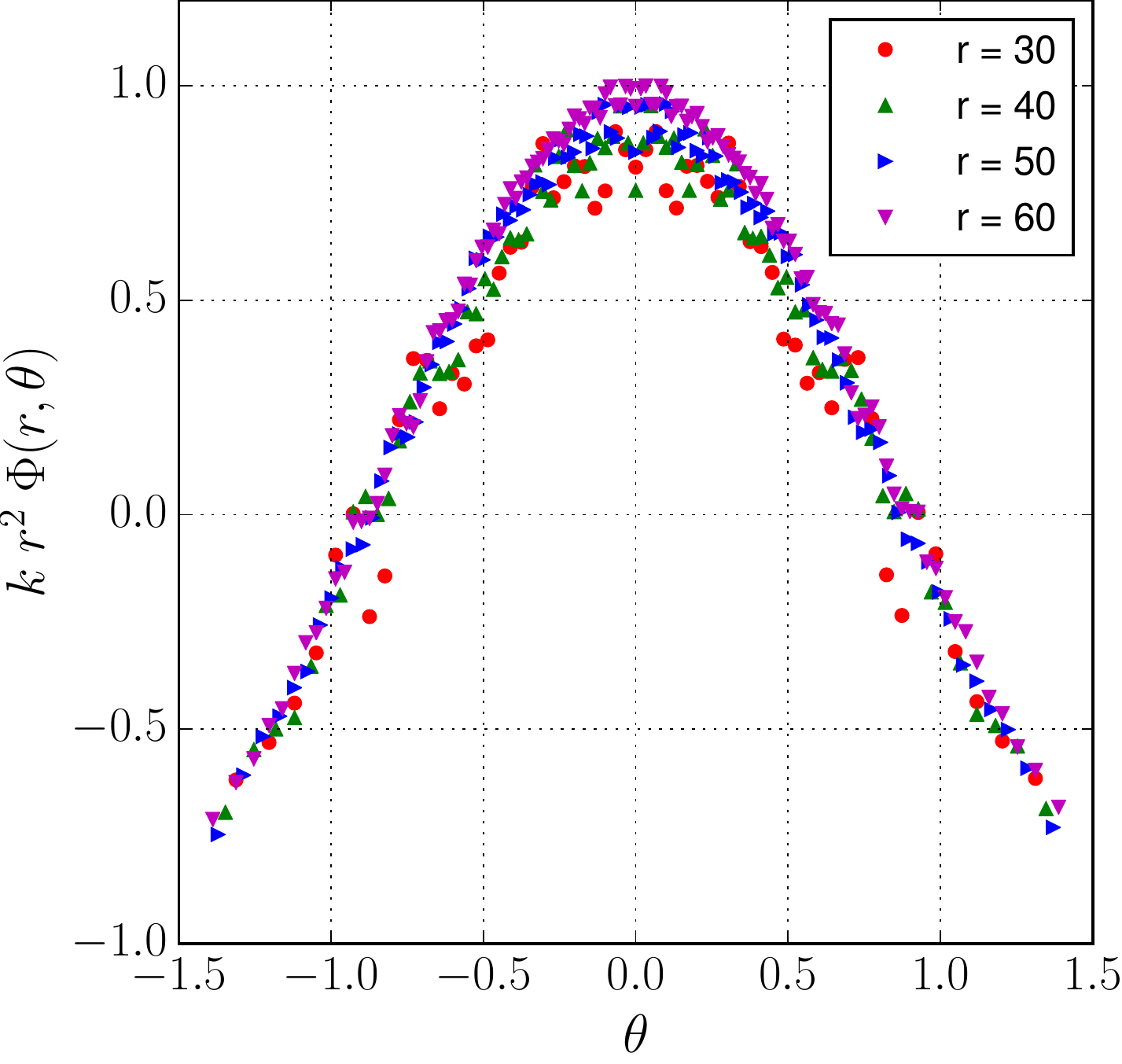}
        %\caption[d]{}  
    \end{subfigure}
%    \quad
    \begin{subfigure}[b]{0.49\textwidth}
        \centering 
        \includegraphics[width=\textwidth]{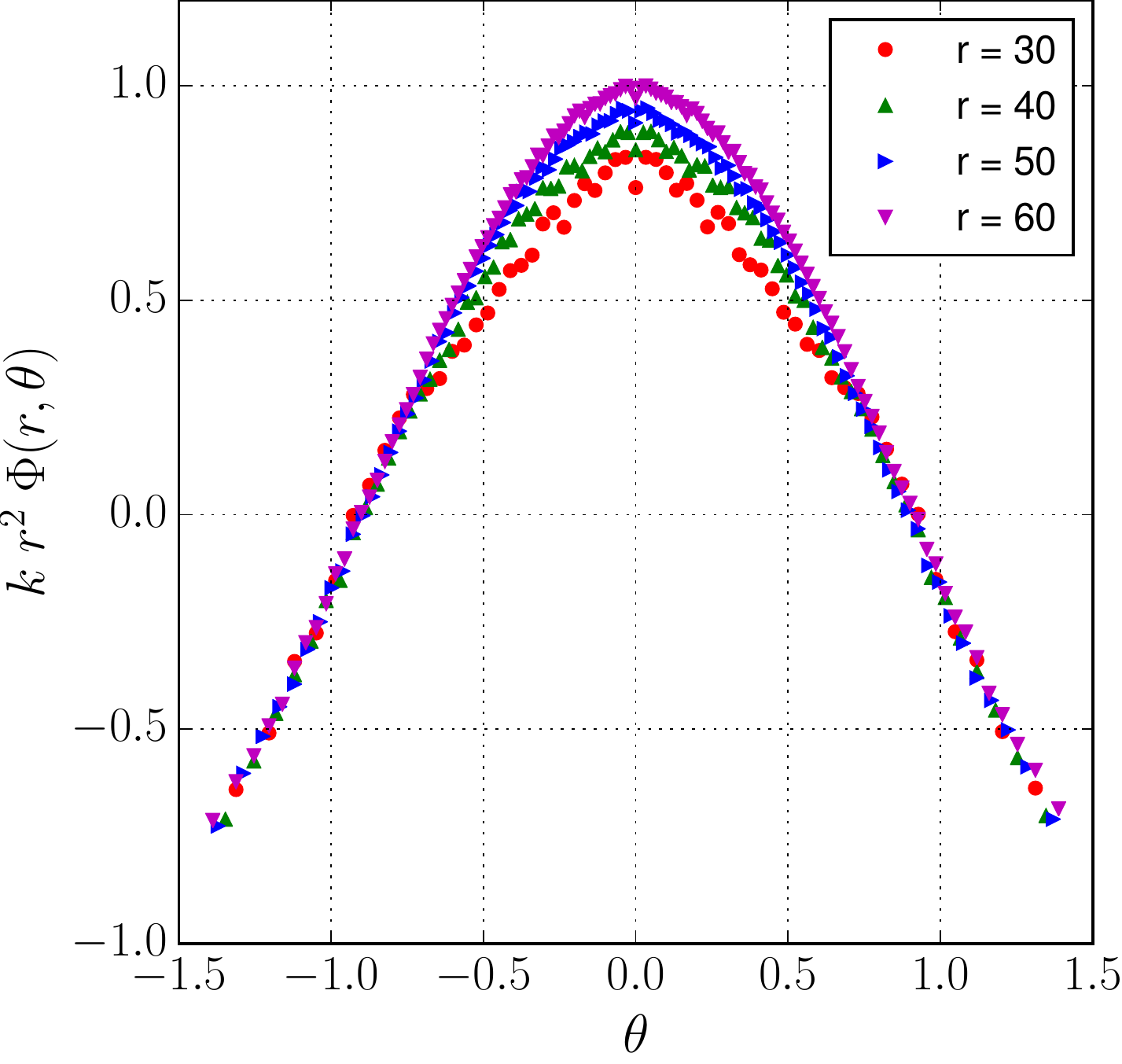}
        %\caption[f]{}  
    \end{subfigure}     
    \caption{ Electric potential measured at several fixed distances $r$ from
              the current injector. Plots present the quantity 
              $r^2~\Phi(r,\theta)$ normalized to $\Phi(40,0)$, showing that
              simulated data points collapse onto a single line, 
              as predicted by Eq.~\ref{eq:phi-approx-inlet}.
              Results taken from a simulation on a $2000 \times 500$ lattice, 
              with an initial uniform density $n = 1.5$, $T = 1.25$, $C_g=10$
              and a fixed velocity at inlet $v^{in} = 10^{-5}$ (all quantities in dimensionless units). 
              Left: $\tau=0.8$. Right: $\tau=1.2$.
    }\label{fig:phi-inlet}
\end{figure}

Using this setup, we simulate a system similar to the one considered in 
\cite{torre-2015,bandurin-2016}, where analytical results are obtained in the approximation of
an infinitely long channel; we use a lattice with an aspect ratio $L/W = 4$, that we simulate
on a lattice of $2000 \times 500$ grid points. The translation between physical units and adimensional
lattice units is based on the definition of a length-unit on the lattice such that the width of the
channel corresponds to the physical value and on an energy unit that we chose as the Fermi-energy
of the simulated system. In \autoref{fig:whirlpool-rlbm} we show a snapshot of a simulation,
using a constant initial density and a large value for the shear viscosity. 
%In numerical units $n=1, T=1, \tau = 1, \eta = \frac{3}{8}$, 
%the velocity at the inlet $v^{in} = 10^{-5} $.
As we can see, results are qualitatively comparable with those presented in
\cite{torre-2015,bandurin-2016}. In particular one can appreciate the (symmetric)  
formation of electron back-flows in the proximity of the gates, so called current whirlpools.
For a more quantitative comparison, we take into consideration the electrochemical potential 
in the proximity of the injector, for which the following approximate
analytic expression was derived in \cite{torre-2015}:
\begin{equation}\label{eq:phi-approx-inlet}
  \Phi(r, \theta) \approx \frac{2 I \eta}{ \pi \bar{n}^2 e^2} \frac{cos(2 \theta)}{r^2} \quad ;
\end{equation}
$I$ is the driving current at the inlet, $\eta$ is the shear viscosity, $\bar{n}$ is the equilibrium 
density, $e$ is the electron charge, $r$ and $\theta$ are used to parametrize in polar coordinates 
the space in the proximity of the inlet.
In \autoref{fig:phi-inlet} we compare the prediction of Eq.~\ref{eq:phi-approx-inlet} with 
the results of our simulations by plotting the electrochemical potential as a function of the polar angle 
for several lattice points at several distances $r$ from the center of the injector. 
In particular we show that for different setups, the quantity $r^2~\phi(r, \theta)$
does not depend on $r$ as predicted by \autoref{eq:phi-approx-inlet}: 
to a good approximation, all curves collapse on the top of each other, as expected. 

%
% \begin{figure}[H]
%   %\includegraphics[width=\linewidth]{{tau1.0n1.5T1.25v1e-05_2000_500}.pdf}
%   \includegraphics[width=\linewidth]{{RLBM_CG1_TAU1.0}.pdf}
  
%   \caption{ Electric potential measured at several fixed distances $r$ from
%             the current injector. Left: plot of 
%             $r^2~\phi(r,\theta)$ normalized to $\phi(40,0)$, showing that
%             simulated data points collapse onto a single line, as predicted by 
%             Eq.~\ref{eq:phi-approx-inlet}. Right: direct comparison
%             between the simulated data and the analytic prediction for 
%             the potential in the proximity of the injector (Eq.~\ref{eq:phi-approx-inlet}).
%             Results taken from a simulation on a $2000 \times 500$ lattice, 
%             with an initial uniform density $n = 1.5$, $T = 1.25$, 
%             and a fixed velocity at inlet $v^{in} = 10^{-5}$ (all quantities in adimensional lattice units).
%           }\label{fig:phi-inlet}
% \end{figure}

\begin{figure}[htb]
    \centering
    \begin{subfigure}[b]{0.325\textwidth}
        \centering
        \includegraphics[width=\textwidth]{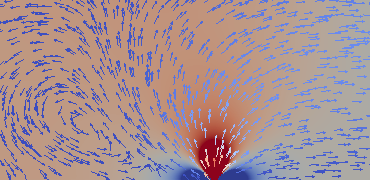}
        \caption[a]{$C_g = 10$}
    \end{subfigure}
%    \quad
    \begin{subfigure}[b]{0.32\textwidth}  
        \centering 
        \includegraphics[width=\textwidth]{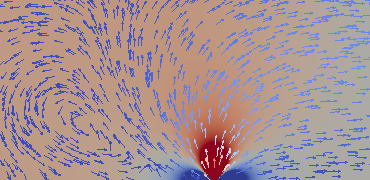}
        \caption[b]{$C_g = 10^4$} 
    \end{subfigure}
%    \quad
    \begin{subfigure}[b]{0.32\textwidth}  
        \centering 
        \includegraphics[width=\textwidth]{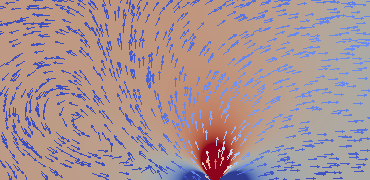}
        \caption[c]{$C_g = \infty$}   
    \end{subfigure}
    \hfill   
    \begin{subfigure}[b]{0.99\textwidth}   
        \centering 
        \includegraphics[width=\textwidth]{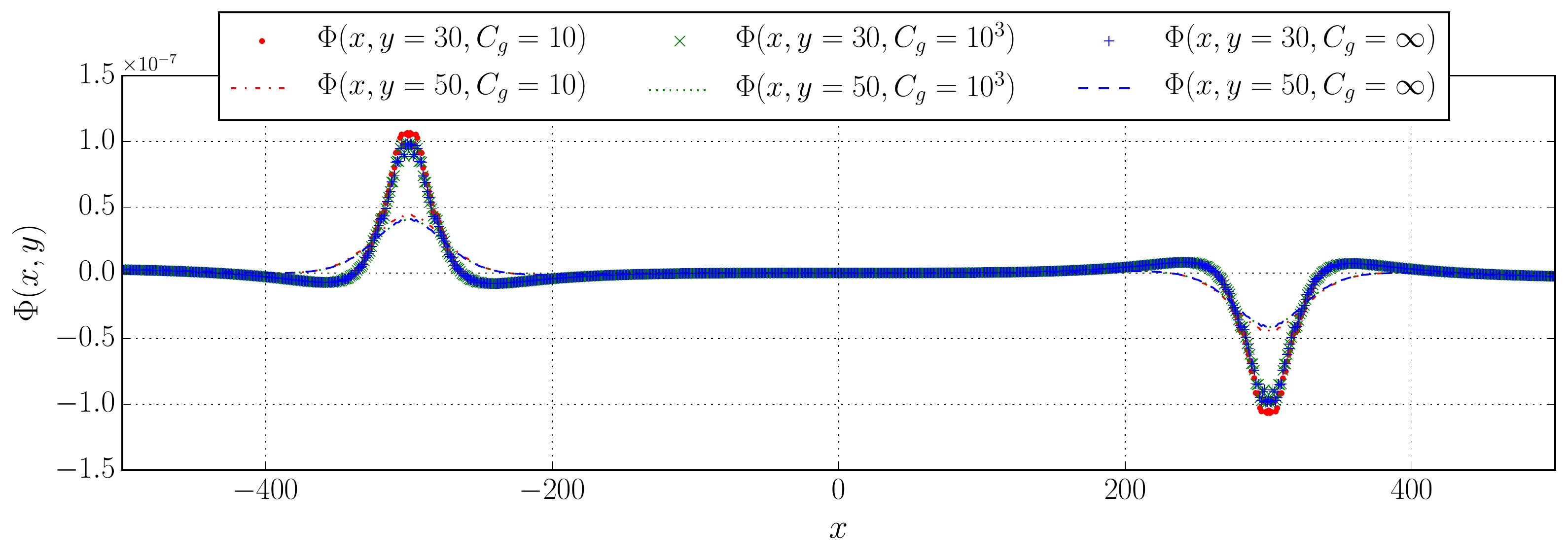}
        \caption[g]{}  
    \end{subfigure}        
    \caption{Qualitative (top) and quantitative (bottom) comparison of 
             the electrochemical potential obtained by varying the intensity of the
             electric field $F_E$. Results taken from a simulation on a $2000 \times 500$ lattice, 
             with an initial uniform density $n = 1.5$, $T = 1.25$, $\tau = 1.0$
             and a fixed velocity at inlet $v^{in} = 10^{-5}$ (all quantities in dimensionless units). } 
    \label{fig:phi-vs-cg}
\end{figure}

As a further benchmark we evaluate how the steady state solution reported in \autoref{fig:whirlpool-rlbm}
varies when tuning the magnitude of the driving forces $F_E$ and $F_P$. To this purpose, we perform simulations with different
values of the parameter $C_g$ (see Eq.~\ref{eq:lca}) to evaluate the role of the electric potential. 
Following \cite{lucas-2016} one would not expect to observe the effect of Coulomb interactions for static flows.
In \autoref{fig:phi-vs-cg} we show that this is indeed the case; in fact varying $C_g$ over several different orders 
of magnitude does not yield any appreciable effect on the solution. Moreover the results
are the same even in the case when $F_E$ is neglected ($C_G = \infty$), proving that the model gives a 
self-consistent description of hydrodynamic theory on long length scales. On the other hand, the electric potential is expected to play a major role on the dynamics of non-linear, time-dependent flows, which will make the object of forthcoming studies.

%

%=====================================================================================
\section{Conclusions}
%=====================================================================================

In this work we have described a new solver for the study of (2+1)-dimensional 
relativistic hydrodynamics based on the Lattice Boltzmann Method. 
The model is applied to the specific study case of the analysis of the electrons flow in graphene.
We have presented results of simulations of a doped single graphene layer  
sheet in the so-called "vicinity-geometry". From a qualitative point of view
we have successfully reproduced the current whirlpools highlighted by recent
experimental works \cite{bandurin-2016}. Besides, we have provided a more
quantitative  validation, with a comparison of the electrochemical potential in the
proximity of the current  injector against previous analytic predictions
\cite{torre-2015}.
We consider this to be a first step in the derivation of an accurate model for the study
of the hydrodynamics behavior of electrons flow in graphene. Future works will 
deal with more robust comparisons of simulations against experimental data and more 
detailed simulations of actual experimental setups.
This work will allow a proper evaluation of the loss of accuracy due to the 
neglect of quantum effects, alongside with further parameters that
should be taken into account (such as electrons collisions with impurities and phonons) 
to expand the capabilities of the present model.

%=====================================================================================
\section*{Acknowledgements}
%=====================================================================================
The authors would like to thank Marco Polini, Jacopo Torre and Francesco
Pellegrino for fruitful discussions, and Matthias Ehrhardt and Andreas Bartel
for carefully revising the manuscript. 
AG has been supported by the European Union's Horizon 2020 research and
innovation programme under the Marie Sklodowska-Curie grant agreement No. 642069 (HPC-LEAP).
MM and SS thank the European Research Council (ERC) Advanced Grant No.
319968-FlowCCS for financial support. All numerical work has been performed on
the COKA computing cluster  at Universit\`a di Ferrara.  
%=====================================================================================

%===================================================================================================
\newpage
\appendix
\renewcommand{\thesection}{A}
\section{Third order 2D Relativistic Orthonormal Polynomials}\label{sec:appendixA}
%===================================================================================================
%
In this appendix we provide the analytic expressions of the relativistic orthogonal polynomials
for the ultra-relativistic case up to the third order. The notation $J^{(n)}_{m_1 \dots m_n}$,
$m_i \in {0,x,y}$ is used to label the polynomial of order $n$ with the subscript $\mu$ 
referring to the corresponding element of the generating basis 
$ \mathcal{V} = \{ 1, p^{\alpha}, p^{\alpha} p^{\beta} \dots \}$ ($\alpha, \beta \in \{ 0, x, y\}$):
\begin{align*}
  J^{(0)}       &= 1 \\
  J^{(1)}_0     &= p^0-1 \\
  J^{(1)}_x     &= p^x \\
  J^{(1)}_y     &= p^y \\
  J^{(2)}_{00}  &= \frac{1}{2}(p^0)^2-2 p^0+1 \\
  J^{(2)}_{0x}  &= \frac{1}{\sqrt{3}}p^0 p^x-\sqrt{3} p^x \\
  J^{(2)}_{0y}  &= \frac{1}{\sqrt{3}}p^0 p^y-\sqrt{3} p^y \\
  J^{(2)}_{xx}  &= \frac{1}{\sqrt{3}} (p^x)^2 -\frac{1}{2 \sqrt{3}} (p^0)^2 \\
  J^{(2)}_{xy}  &= \frac{1}{\sqrt{3}} p^x p^y \\
  J^{(3)}_{000} &= \frac{1}{6}(p^0)^3 -\frac{3}{2}(p^0)^2 +3 p^0-1 \\
  J^{(3)}_{xxx} &= -p^0 p^x+\frac{1}{6} (p^x)^3 +\frac{3}{2} p^x \\
  J^{(3)}_{00x} &= \frac{1}{\sqrt{15}}(p^0)^2 p^x-\sqrt{\frac{5}{3}} p^0 p^x-\frac{1}{2 \sqrt{15}}(p^x)^3  +\frac{\sqrt{15}}{2}p^x  \\
  J^{(3)}_{0xx} &= -\frac{1}{2 \sqrt{15}}(p^0)^3 +\frac{1}{2} \sqrt{\frac{5}{3}} (p^0)^2+\frac{1}{\sqrt{15}}p^0 (p^x)^2-\sqrt{\frac{5}{3}} (p^x)^2 \\
  J^{(3)}_{00y} &= \frac{1}{2 \sqrt{6}}(p^0)^2 p^y  -2 \sqrt{\frac{2}{3}} p^0 p^y+\sqrt{6} p^y \\
  J^{(3)}_{xxy} &= \frac{1}{3} \sqrt{\frac{2}{5}} (p^x)^2 p^y-\frac{1}{6 \sqrt{10}} (p^0)^2 p^y \\
  J^{(3)}_{0xy} &= \frac{1}{\sqrt{15}}p^0 p^x p^y-\sqrt{\frac{5}{3}} p^x p^y
\end{align*}

%=====================================================================================

%=====================================================================================
%\newpage
\appendix
\renewcommand{\thesection}{B}
\section{Third order 2D Orthogonal Projections}\label{sec:appendixB}
%=====================================================================================

In this appendix we provide the analytic expressions of the orthogonal projections $a^{(k)}$, 
up to the third order, for the ultra-relativistic case. 
The notation follows the one introduced in \autoref{sec:appendixA} for the orthogonal polynomials.
All the projections are scaled with the particle density $n$, thereby ensuring the correct normalization 
of the equilibrium distribution function (Eq.~\ref{eq:maxwell-juttner}).

\begin{align*}
  a^{(0)}       &= 1 \\
  a^{(1)}_0     &= T u^0-1 \\ 
  a^{(1)}_x     &= T u^x \\ 
  a^{(1)}_y     &= T u^y \\ 
  a^{(2)}_{00}  &= \frac{1}{2} T^2 \left(3 (u^0)^2-1\right)-2 T u^0+1 \\ 
  a^{(2)}_{0x}  &= \sqrt{3} T u^x (T u^0-1) \\ 
  a^{(2)}_{0y}  &= \sqrt{3} T u^y (T u^0-1) \\ 
  a^{(2)}_{xx}  &= -\frac{1}{2} \sqrt{3} T^2 \left((u^0)^2-2 (u^x)^2-1\right) \\ 
  a^{(2)}_{xy}  &= \sqrt{3} T^2 u^x u^y \\ 
  a^{(3)}_{000} &= \frac{1}{2} (T u^0-1) \left(T^2 \left(5 (u^0)^2-3\right)-4 T (u^0)+2\right) \\ 
  a^{(3)}_{xxx} &= \frac{1}{2} T u^x \left(T^2 \left(5 (u^x)^2+3\right)-6 T u^0+3\right) \\ 
  a^{(3)}_{00x} &= -\frac{1}{2} \sqrt{15} T u^x \left(T^2 \left(-2 (u^0)^2+(u^x)^2+1\right)+2 T u^0-1\right) \\ 
  a^{(3)}_{0xx} &= -\frac{1}{2} \sqrt{15} T^2 (T u^0-1) \left((u^0)^2-2 (u^x)^2-1\right) \\ 
  a^{(3)}_{00y} &= \frac{1}{2} \sqrt{\frac{3}{2}} T u^y \left(T^2 \left(5 (u^0)^2-1\right)-8 T u^0+4\right) \\ 
  a^{(3)}_{xxy} &= -\frac{1}{2} \sqrt{\frac{5}{2}} T^3 u^y \left((u^0)^2-4 (u^x)^2-1\right) \\ 
  a^{(3)}_{0xy} &= \sqrt{15} T^2 u^x u^y (T u^0-1)
\end{align*}

%=====================================================================================

%=====================================================================================
\newpage
\appendix
\renewcommand{\thesection}{C}
\section{Third order expansion of the equilibrium distribution function}\label{sec:appendixC}
%=====================================================================================

The third order expansion of the Maxwell-J\"uttner distribution in two dimension and for $m=0$,
which allows to recover the first, the second and the third order moments of Eq.~\ref{eq:maxwell-juttner},
was derived using the polynomials defined in \autoref{sec:appendixA} and projections in 
\autoref{sec:appendixB}. It reads as follow:

\begin{align*}
 f_i^{eq} &= \frac{w_i~n}{T} \left(\frac{1}{6} T^3 u^x \left( (p_i^x)^3-\frac{3 (p_i^0)^2 p_i^x}{4}\right) \left(-3 (u^0)^2+4 (u^x)^2+3\right) \right. \\
          & \left. +\frac{1}{24} p_i^y T^3 u^y \left((p_i^0)^2 -4 (p_i^x)^2\right) \left((u^0)^2-4 (u^x)^2-1\right) \right. \\
          & \left. +\frac{1}{4} T^2 \left((p_i^0)^2-2 (p_i^x)^2\right) \left((u^0)^2-2 (u^x)^2-1\right)+\frac{1}{8} \left(-2 (p_i^0)^2+(p_i^0-5) (p_i^x)^2 \right. \right.\\
          & \left. \left. +9 p_i^0-3\right) (T u^0-1) \left(T \left(5 T (u^x)^2+T-2 u^0\right)+1\right)+\frac{1}{24} \left(4 (p_i^0)^3-30 (p_i^0)^2 \right. \right.\\
          & \left. \left. -3 (p_i^0-5) (p_i^x)^2+45 p_i^0-15\right) (T u^0-1) \left(T^2 \left(4 (u^0)^2-3 \left((u^x)^2+1\right)\right)  \right. \right.\\
          & \left. \left. -2 T u^0+1\right)+(p_i^0-5) p_i^x p_i^y T^2 u^x u^y (T u^0-1) \right.\\
          & \left. +\frac{1}{8} (p_i^0-6) (p_i^0-2) p_i^x T u^x \left(T^2 \left(5 (u^0)^2-1\right)-8 T u^0+4\right) \right.\\
          & \left. +(p_i^0-3) p_i^x T u^x (T u^0-1)+\frac{1}{8} (p_i^0-6) (p_i^0-2) p_i^y T (u^y) \left(T^2 \left(5 (u^0)^2-1\right) \right. \right.\\
          & \left. \left. -8 T u^0+4\right)+(p_i^0-3) p_i^y T u^y (T u^0-1)+\frac{1}{4} ((p_i^0-4) p_i^0+2) \left(T^2 \left(3 (u^0)^2 \right. \right. \right.\\
          & \left. \left. \left. -1\right)-4 T u^0+2\right)+(p_i^0-1) (T u^0-1)+p_i^x p_i^y T^2 u^x u^y+p_i^x T u^x+p_i^y T u^y+1\right) \\
\end{align*}

%=====================================================================================

%=====================================================================================
\section*{References}
\bibliography{references}
%=====================================================================================

%=====================================================================================
\end{document}